\RequirePackage{fix-cm}
\documentclass[smallcondensed,final]{svjour3}  
\smartqed  % flush right qed marks, e.g. at end of proof
\usepackage{cite}   
\usepackage{amsmath}
\usepackage{hyperref}
\hypersetup{
     colorlinks   = true,
     citecolor    = blue
}
\usepackage{graphicx}
\usepackage{braket}
\usepackage{bbold}
\usepackage[round]{natbib}   % omit 'round' option if you prefer square brackets
\graphicspath{{images/}}
\begin{document}
\title{Non-Markovian resonance fluorescence}


\author{Abhishek Kumar}


\institute{Abhishek Kumar \at
              Reykjavik University, \\
              School of Science and Engineering, \\
              Menntavegur 1, IS-101 Reykjavik, Iceland \\
              Tel.: +91-9987868469\\
              \email{srivastvaster@outlook.com}           
}

\date{Received: date / Accepted: date}

\maketitle

\begin{abstract}
We derive a general formula for the non-Markovian fluorescence spectrum of a multi-level system interacting with a bosonic environment. To this end, we apply linear-response theory to describe the dynamics of a detector monitoring the emission spectrum of a general multi-level system.  The resultant emission lineshape function is directly related to the two-time correlation of system observables, which we derive using Nakajima-Zwanzig Generalized Master Equation without assuming a Markov approximation.
\end{abstract}

%%insert keywords separated by comma using \keywords{words}
%%include \pacs{number} to print the PACS number
%\pacs{12.60.Jv; 12.10.Dm; 98.80.Cq; 11.30.Hv}}]
%%close the twocolumn escape here

%%include \doinum{number}for the DOI number in the header
%%include \volnum{number} for the volume number in the header
%%include \year{yyyy} for  year of publication in the header
%%include \pgrange{num--num} page range of article in the header
%%include \artcitid{num} for the article citation id
%%include \lp to print last page of the article
%%include \setcounter{page}{pagenum} for the exact starting page of the article

%\doinum{12.3456/s78910-011-012-3}
%\artcitid{\#\#\#\#}
%\volnum{123}
%\year{2016}
%\pgrange{23--25}
%\setcounter{page}{23}
%\lp{25}
\section{Introduction} \label{sec:one}
%% If no rules required then use the following:
%% \begin{strip}
%% wide matter (a formula) in two-column mode
%% \end{strip}
Resonance fluorescence of a multi-level system, strongly driven by a laser field, has been an active area of research both theoretically and experimentally \citep{peng2019, Konthasinghe2019, scholl2019, yang2016, ulrich11prl}. Moreover, most of the theoretical work to compute the resonance fluorescence in the context of open quantum system, where system $\mathbf{S}$ interacts with its environment $\mathbf{E}$, has been developed in terms of a purely Markovian decay process. A Markovian decay process assumes that the correlation time is much smaller than the system decay time and it is not affected by the time scales close to system decay time. Once the theory has been relied on the assumption of a Markovian (history independent) decay process, the quantum regression theorem (QRT) can be applied to calculate the dynamics of system correlation function \citep{mollow69,swan81,lax00,lax68}. However, the dominant real physical processes in such interactions (e.g. nuclear spins, phonons) \citep{bill04,abhi2018,krummheuer02,vagov14,weiler12,hughes12prb} are known to follow a non-Markovian (history-dependent) dynamics. Therefore, the fluorescence lineshape predicted using a Markovian theory and applying QRT can overlook the correct physical process in the interaction and lead to incorrect results.

In this paper, we intend to perform a detailed theoretical analysis to compute a general formula for resonance fluorescence spectrum of a multi-level system undergoing a non-Markovian decay process. We have used the Nakajima-Zwanzig generalized master eqution (GME) \citep{fick} to calculate the dynamics of system's two-time correlation function to all orders beyond Markov approximation, which is directly related to the resultant fluorescence spectrum \citep{mollow69,vega08}. Moreover, we discuss the presence of an additional non-vanishing term in the two-time correlation known as \emph{final term} which does not allow us to describe the dynamics within QRT, since this will essentially lead to applying a Markovian approximation \citep{guarnieri2014,ford96}. This \emph{final term} can be neglected under Markovian assumption and then the well-known \emph{quantum regression theorem} can be applied to compute the system correlation function \citep{swan81}.

In order to illustrate the theory, we have already discussed the fluorescence spectra of a laser driven two-level system, embedded in a cavity and coupled to a three-dimensional bath of acoustic phonons within non-Markovian regime in our previous paper \citep{abhi2018}.

The structure of this paper is the following. In Sec. \ref{sec:two}, we present a general Hamiltonian of a multi-level system $\mathbf{S}$ which we want to discuss along with its environment $\mathbf{R}$. In Sec. \ref{sec:three}, we discuss a theoretical setup and its general Hamiltonian in terms of system and environment observables. In Sec. \ref{sec:four}, we discuss the linear response theory to describe the dynamics of emitted radiation in terms of system's observables. In Sec. \ref{sec:five}, we discuss and derive the Nakajima-Zwanzig GME to compute the two-time correlation and finally obtain the expression for fluorescence lineshape function. Sections \ref{sec:six} and \ref{sec:seven} are devoted to discuss the results and conclusion, respectively.
\section{Model Hamiltonian}\label{sec:two}

We start with the Hamiltonian of a general multi-level system interacting with radiation modes of the electromagnetic field, which can be written as system ($H_S$), field ($H_R$), and interaction ($H_{SR}$) in terms of the standard Jaynes-Cummings model within a rotating-wave approximation,
%(see fig.\ $\ref{fig:gedanken setup for multilevel}$):
\begin{align}
H&=H_{0}+H_{SR}\\ \label{eqn:emitting multi level system Hamiltonian}
H_{0}&=H_{S}+H_{R}\\
H_{R}&=\sum_{k}\hbar\omega_{k}a_{k}^{\dagger}a_{k}\\
H_{SR}&=\sum_{k}\hbar g_{k}(\sigma_{+}a_{k}+\sigma_{-}a_{k}^{\dagger}),\label{eqn:coupling constant for system}
\end{align}
where  $\sigma_{+}=|a\rangle\langle b|$ and $\sigma_{-}=|b\rangle\langle a|$ are the raising and lowering operators between a selected excited state $|a\rangle$ and ground state $|b\rangle$ in the Hilbert space of the system and $a_{k}$, $a_{k}^{\dagger}$ are the annihilation and creation operators in the Hilbert space of a set of electromagnetic modes coupled to the system. The coupling to the radiation modes is given by the coupling constant $g_{k}$ to a mode of frequency $\omega_{k}$. The coupling constant $g_{k}$ is related to the electric-dipole transition matrix element $|\rho_{ab}|$ through Eq.\ $(\ref{eqn:coupling constant for system})$.
\section{Fluorescence spectra}\label{sec:three}
We theoretically describe a realistic detector with the well-known gedanken spectrum analyzer \citep{scully, cohen} which measures the scattered fluorescence light from the emitting system.
We assume that the radiation field emitted by the system is detected by a two-level (detector) atom, with transition frequency $\omega_{\alpha}-\omega_{\beta}=\omega_{0}$. The detector atom has sharp levels $|\alpha\rangle$ and $|\beta\rangle$ separated by energy $\omega_{0}$ and is initially prepared in the ground state $|\beta\rangle$ and is placed inside a shutter, that only opens during a certain observation time $T$, during which it receives the emitted radiation and may be excited to the upper level $|\alpha\rangle$.  The excitation rate of the detector sets the detector response profile, centered at frequency $\omega_{0}$ with bandwidth $\Delta\omega_{B}$. The detector bandwidth is assumed to be small compared to the typical feature size in the fluorescence spectrum. The detector atom has a Hamiltonian of the form,
\begin{equation}
H_{D}=\frac{\hbar\omega_{0}}{2}\left(|\alpha\rangle\langle\alpha|-|\beta\rangle\langle\beta|\right).
\end{equation}
The total Hamiltonian (for system, field and detector) has the form,
\begin{equation}
H_{T}=H+H_{D}+H_{DR}.
\label{eqn:total Hamiltonian system, field and detector}
\end{equation}
The coupling between detector atom $H_{D}$ and the radiation field, included in the Hamiltonian $H$, is given by the Hamiltonian 
\begin{equation}
H_{DR}=\sum_{k}\hbar g_{k}^{D}\left(|\alpha\rangle\langle\beta|a_{k}+|\beta\rangle\langle\alpha|a_{k}^{\dagger}+|\alpha\rangle\langle\beta|a_{k}^{\dagger}+|\beta\rangle\langle\alpha|a_{k}\right),
\end{equation}
where $g_{k}^{D}$ is the coupling of the detector to field mode $k$ and the detector coupling Hamiltonian $H_{DR}$, in the interaction picture with respect to the Hamiltonian $H_{0}'=H+H_{D}$, is given by
\begin{equation}
H_{DR}^{I}(t)=e^{iH_{0}'t/\hbar}\,H_{DR}\,e^{-iH_{0}'t/\hbar},
\end{equation}
we adopt the rotating wave approximation which is equivalent to dropping the energy non-conserving terms, to write
\begin{equation}
H_{DR}^{I}(t)=\sum_{k}\hbar\,g_{k}^{D}\left(\sigma_{\alpha\beta}\,a_{k}(t)\,e^{i\omega_{0}t}
+\sigma_{\beta\alpha}\,a_{k}^{\dagger}(t)\,e^{-i\omega_{0}t}\right),
\label{eqn:detector coupling hamiltonian in ip}
\end{equation}
Here, $\sigma_{\alpha\beta}=|\alpha\rangle\langle\beta|$, $\sigma_{\beta\alpha}=|\beta\rangle\langle\alpha|$ and $a_{k}(t)$ is in the interaction picture with $H_{0}'$, and we assume that the electric field is linearly polarized along the $x$-axis. The positive-frequency part of the electric field is defined by
\begin{equation}
E^{+}(t)=\sum_{k}\varepsilon_{k}a_{k}(t),
\end{equation} 
the negative-frequency part of the electric field is $E^{-}(t)=\left[E^{+}(t)\right]^{\dagger}$ and the detector coupling $g_{k}^{D}$ is related to the electric-dipole transition matrix element $|\rho_{\alpha\beta}^{k}|$ through \citep{scully},
\begin{equation}
g_{k}^{D}=\frac{|\rho_{\alpha\beta}^{k}|\varepsilon_{k}}{\hbar}.
\label{eqn:coupling constant for detector}
\end{equation}

In the long-wavelength limit, we take the dipole element $\rho_{\alpha\beta}\approx\rho_{\alpha\beta}^{k}$ (independent of $k$). From Eqs. (\ref{eqn:detector coupling hamiltonian in ip}) and (\ref{eqn:coupling constant for detector}), we find the detector coupling Hamiltonian
\begin{equation}
H_{DR}^{I}(t)=\rho_{\alpha\beta}\,\sigma_{\alpha\beta}\,E^{+}(t)\,e^{i\omega_{0}t}
+\rho_{\alpha\beta}\,\sigma_{\beta\alpha}\,E^{-}(t)\,e^{-i\omega_{0}t},
\label{eqn:detector coupling Hamiltonian in interaction picture}
\end{equation}
where $\rho_{\alpha\beta}=e\langle\alpha|x|\beta\rangle=\rho_{\alpha\beta}^{*}$. From here and what follows, for the simplicity we will use $\hbar=1$.
\section{Linear response theory}\label{sec:four}
Linear response theory \citep{bruus} states that the response to a weak external perturbation is proportional to the perturbation, and therefore all we need to understand is the proportionality constant. We consider that the detector is weakly coupled to the scattered emitted field, which allows us to do the linear response on the detector coupling. With this setup, the idea is to calculate the spectrum of light emitted by the system in the presence of a detector coupled to the emitted field at some point in time $t>0$. This spectrum is defined in terms of the probability to excite the detector atom, $P(\omega_{0},t)$ at time $t$. 
The probability of exciting the detector atom to excited level $|\alpha\rangle$
is found by calculating the expectation value of the projection
operator $|\alpha\rangle\langle\alpha|$,which can be evaluated as
\begin{equation}
P(\omega_{0},t)=\langle\psi_{I}(t)|\alpha\rangle\langle\alpha|\psi_{I}(t)\rangle,
\label{eqn:probability of being in upper state}
\end{equation}	
where $|\psi(t)\rangle$ is the interaction-picture state, and is given by
\begin{equation}
|\psi_{I}(T)\rangle=U_{I}(T)|\psi_{I}(0)\rangle
\end{equation}
$U_{I}(T)$ is the time-evolution operator given by
\begin{align}
U_{I}(T)&=\mathcal{T}\mathrm{exp}\lbrace-i\int_{0}^{T}dtH_{DR}^{I}(t)\rbrace\\
&=1-i\int_{0}^{T}dt_{1}H_{DR}^{I}(t_{1})-\frac{1}{2}\mathcal{T}\int_{0}^{T}dt_{1}\int_{0}^{T}dt_{2}H_{DR}^{I}(t_{1})H_{DR}^{I}(t_{2})+.....
\end{align}
We assume that the detector atom is initially in the ground state $|\beta\rangle$ and that the initial state of the system, radiation modes, and detector is given by the product state $|\psi_{SR}(0)\rangle\otimes|\beta\rangle$, where $|\psi_{SR}(0)$ is an eigenstate of the Hamiltonian $H$, including the coupling between system and the radiation modes.
Suppose now that at some time t=0, an external perturbation is applied, driving the system out of equilibrium. The perturbation is described by the term $H_{DR}^{I}(t)$. Now we wish to find the expectation value of the operator $|\alpha\rangle\langle\alpha|$ at time $T>0$. In order to do so, we must find the time evolution of the state $|\psi_{I}(T)\rangle$ in the interaction picture with the interaction picture Hamiltonian $H_{DR}^{I}(t)$. To linear order in $H_{DR}^{I}$, we obtain an expression for the state in the interaction picture up to first order in the perturbation

\begin{equation}
|\psi_{I}(T)\rangle \approx \left[1-i\int_{0}^{T}dt_{1}H_{DR}^{I}(t_{1})\right]|\psi_{SR}(0)\rangle.
\label{eqn:linear response theory}
\end{equation}
Going back to the Schr\"odinger picture, we have
\begin{equation}
|\psi(T)\rangle =e^{-iH_{0}'t} \left[1-i\int_{0}^{T}dt_{1}H_{DR}^{I}(t_{1})\right]|\psi_{SR}(0)\rangle.
\end{equation}
Therefore the excitation probability, in the Schr\"odinger picture, takes the form
\begin{align}
P(\omega_{0},t=T)&=\langle\psi(0)|U_{I}^{\dagger}(T)e^{iH_{0}'T}|\alpha\rangle\langle\alpha|e^{-iH_{0}'T}U_{I}(T)|\psi(0)\rangle\nonumber\\
&=\langle\psi(0)|U_{I}^{\dagger}(T)|\alpha\rangle\langle\alpha|U_{I}(T)|\psi(0)\rangle,
\end{align}	
here we have used $[H_{0}',|\alpha\rangle\langle\alpha|]=0$. By substituting the values of $H_{DR}^{I}(t_{1})$ and then $|\psi(T)\rangle$ in equations (\ref{eqn:linear response theory}) and (\ref{eqn:probability of being in upper state}), respectively, the resulting expression for the
excitation probability in the interaction picture is calculated as,
\begin{equation}
P(\omega_{0},T)=|\rho_{\alpha\beta}|^2\int_{0}^{T}dt_{1}\int_{0}^{T}dt_{2}\langle E^{-}(t_{1})
E^{+}(t_{2})\rangle e^{-i\omega_{0}(t_{1}-t_{2})}.
\label{eqn:excitation probability in terms of positive and negative electric field operators}
\end{equation}
Here, we have generalized to a mixed-state initial condition and the average $\langle ...\rangle=\mathrm{Tr}\lbrace ...\rho(0)\rbrace$, where $\rho(0)$
is the total density matrix of the scattering system plus detector and the radiation field at time $T=0$ is \emph{not} a product state. From Eq.\ (\ref{eqn:excitation probability in terms of positive and negative electric field operators}), it follows that the excitation probability, $P(\omega_{0},T)$, of the detector atom is proportional to the two-time correlation function of the field operators, i.e., $\langle E^{-}(t_{1})E^{+}(t_{2})\rangle$. We write the probability of excitation in terms of observables of the emitting system. We start with the Heisenberg equation of motion for $a_{k}(t)$ with the Hamiltonian $H$ given by Eq. (\ref{eqn:emitting multi level system Hamiltonian}), i.e.,
\begin{equation}
\frac{da_{k}(t)}{dt}=-i[a_{k}(t),H],
\end{equation}
we obtain an equation for $a_{k}(t)$
\begin{equation}
\dot{a}_{k}(t)=-i\biggl(\omega_{k}a_{k}(t)+g_{k}\sigma_{-}(t)\biggr) .
\end{equation}
After integrating the above equation starting from time $t=t_{0}<0$, we obtain the expression
\begin{equation}
a_{k}(t)=a_{k}(0)e^{-i\omega_{k}t}-i\int_{0}^{t}d\tau g_{k}\sigma_{-}
(\tau)e^{-i\omega_{k}(t-\tau)}.
\end{equation}
We insert the above expression for $a_{k}(t)$ into the positive-frequency part of the electric field operator, giving
\begin{equation}
E^{+}(t)=\sum_{k}\varepsilon_{k}a_{k}(0)e^{-i\omega_{k}t}
-i\sum_{k}\varepsilon_{k}g_{k}\times\int_{0}^{t}d\tau\sigma_{-}(\tau)e^{-i\omega_{k}(t-\tau)}.
\label{eqn:positive electric field}
\end{equation}
Similarly, the negative frequency part of the electric field operator is
\begin{equation}
E^{-}(t)=\sum_{k}\varepsilon_{k}a_{k}^{\dagger}(0)e^{i\omega_{k}t}
+i\sum_{k}\varepsilon_{k}g_{k}\times\int_{0}^{t}d\tau\sigma_{+}(\tau)e^{i\omega_{k}(t-\tau)}.
\label{eqn:negative electric field}
\end{equation}
%The first term in Eq.\ (\ref{eqn:positive electric field}) corresponding to the vacuum field, does not contribute to photo detection,
%since the radiation modes are initially in the vacuum state $|0_{k}\rangle$,
%\begin{equation}
%\langle 0_{k}|a_{k}^{\dagger}=0,\,a_{k}|0_{k}\rangle=0. 
%\end{equation}
Replacing equations (\ref{eqn:positive electric field}) and (\ref{eqn:negative electric field}) in eqn. (\ref{eqn:excitation probability in terms of positive and negative electric field operators}), we get (neglecting terms $\sim\langle a_{k}a_{k}^{\dagger}\rangle$, $\langle a_{k}\sigma_{+}\rangle$ and $\sigma_{-}a_{k}^{\dagger}$)
\begin{equation}
P(\omega_{0},T)=\int_{0}^{T}dt_{1}\int_{0}^{T}dt_{2}\,e^{-i\omega_{0}(t_{1}-t_{2})}\times \biggl\lbrace\int_{0}^{t_{1}}d\tau'\int_{0}^{t_{2}}d\tau S^{*}(t_{1}-\tau')
S(t_{2}-\tau)\langle\sigma_{+}(\tau')\sigma_{-}(\tau)\rangle \biggr\rbrace,
\label{eqn:prob of excitation at time T}
\end{equation}
where $S(t-\tau)$ is the detector response function
\begin{align}
S(t-\tau)&=\sum_{k}g_{k}g_{k}^{D}e^{-i\omega_{k}(t-\tau)}\\
&=\alpha\sum_{k}g_{k}^{2}e^{-i\omega_{k}(t-\tau)}
\label{eqn:detector response function}
\end{align}
here we have used $g_{k}^{D}=|\rho_{\alpha\beta}|\varepsilon_{k}$ and $\alpha=\frac{g_{k}^{D}}{g_{k}}$ is the ratio of the detector and system coupling strengths. The detector response function $S(t)$ typically decays rapidly with a decay time $\tau_{\mathrm{cd}}$, which is given by the inverse of the detector bandwidth, $\Delta\omega_{\mathrm{B}}$.
\begin{equation}
\tau_{\mathrm{cd}}=\frac{1}{\Delta\omega_{\mathrm{B}}}
\end{equation}
whereas $\langle\sigma_{+}(\tau')\sigma_{-}(\tau)\rangle$ evolves on a typical time scale $T_{\mathrm{sys}}$, where, by assumption, $T_{\mathrm{sys}}\gg\tau_{\mathrm{cd}}$. We apply change of variables in the Eq.\ $(\ref{eqn:prob of excitation at time T})$ as $\tau'=t_{1}-\tau'$ and $\tau=t_{2}-\tau$ to write
\begin{equation}
P(\omega_{0},T)=\int_{0}^{T}dt_{1}\int_{0}^{T}dt_{2}\,e^{-i\omega_{0}(t_{1}-t_{2})}\times \biggl\lbrace\int_{0}^{t_{1}}d\tau'\int_{0}^{t_{2}}d\tau S^{*}(\tau')
S(\tau)\langle\sigma_{+}(t_{1}-\tau')\sigma_{-}(t_{2}-\tau)\rangle \biggr\rbrace.
\end{equation}
When $\langle\sigma_{+}(t_{1}-\tau')\sigma_{-}(t_{2}-\tau)\rangle$ is a slow varying function where $S(\tau)$ is finite, we approximate: $\langle\sigma_{+}(t_{1}-\tau')\sigma_{-}(t_{2}-\tau)\rangle\approx\langle\sigma_{+}(t_{1})\sigma_{-}(t_{2})\rangle$ and extend the upper limit of integrations to $t_{1,2}\rightarrow\infty$ for $t_{1,2}\gg\tau_{cd}$, we obtain
\begin{equation}
P(\omega_{0},T)\approx|\bar{I}|^2\int_{0}^{T}dt_{1}\int_{0}^{T}dt_{2}\,e^{-i\omega_{0}(t_{1}-t_{2})}\times \biggl\lbrace\langle\sigma_{+}(t_{1})\sigma_{-}(t_{2})\rangle \biggr\rbrace,
\label{eqn:prob in terms of I1 and I2}
\end{equation}
where $\bar{I}=\int_{0}^{\infty}d\tau S(\tau) \propto g\,g^{D}\tau_{cd}$. Rewriting Eq.\ (\ref{eqn:prob in terms of I1 and I2})
\begin{equation}
P(\omega_{0},T)\approx|\bar{I}|^2\int_{0}^{T}dt_{1}\int_{0}^{T}dt_{2}\,e^{-i\omega_{0}(t_{1}-t_{2})}\times \biggl\lbrace\mathrm{Tr}\lbrace\sigma_{-}(t_{2}-t_{1})\rho(t_{1})\sigma_{+}\rbrace \biggr\rbrace,
\end{equation}
here we have used the cyclic property of trace. We apply change of variables in the above equation, $x=t_{2}-t_{1}$, to write
\begin{equation}
P(\omega_{0},T)\approx|\bar{I}|^2\int_{0}^{T}dt_{1}\int_{-t_{1}}^{T-t_{1}}dx\,e^{i\omega_{0}x}\times \biggl\lbrace \mathrm{Tr}\lbrace\sigma_{-}(x)\rho(t_{1})\sigma_{+}\rbrace \biggr\rbrace.
\end{equation}
Here, we will work in the regime where the observation time $T$ is much larger than the system evolution time $T_{\mathrm{sys}}$. In this limit, we take $T-t_{1}\rightarrow\infty$ and $-t_{1}\rightarrow -\infty$, therefore
\begin{equation}
P(\omega_{0},T)\approx|\bar{I}|^{2}\int_{0}^{T}dt_{1}\int_{-\infty}^{\infty}dx\,e^{i\omega_{0}x}\mathrm{Tr}\biggl\lbrace\sigma_{-}(x)\rho(t_{1})\sigma_{+}\biggr\rbrace.
\end{equation}
The contribution of the x-integration from the boundaries is small in the parameter $\frac{T_{\mathrm{sys}}}{T}\rightarrow 0$
\begin{align}
P(\omega_{0},T)&\approx|\bar{I}|^{2}\int_{-\infty}^{\infty}dx\,e^{i\omega_{0}x}\mathrm{Tr}\biggl\lbrace\sigma_{-}(x)\int_{0}^{T}dt_{1}\rho(t_{1})\sigma_{+}\biggr\rbrace\nonumber\\
&=T|\bar{I}|^{2}\int_{-\infty}^{\infty}dx\,e^{i\omega_{0}x}\mathrm{Tr}\biggl\lbrace\sigma_{-}(x)\rho_{T}\sigma_{+}\biggr\rbrace.
\end{align}
where the time-averaged density matrix is
\begin{equation}
\rho_{T}=\frac{1}{T}\int_{0}^{T}dt_{1}\rho(t_{1}).
\end{equation}
\subsection{Fluorescence spectrum in the stationary regime} \label{subsec:one}
The fluorescence spectrum $F(\omega_{0})$ in terms of the two-time correlation function of system observables
and in the stationary limit has the form

\begin{align}
F(\omega_{0})&=\lim_{T\rightarrow\infty}\frac{1}{T}P(\omega_{0},T)\nonumber\\
&=\lim_{T\rightarrow\infty}\frac{1}{T}|\bar{I^{*}}|^{2}\int_{-\infty}^{\infty}dt\,e^{i\omega_{0}t}\mathrm{Tr}\biggl\lbrace\sigma_{-}(t)T\bar{\rho}\sigma_{+}\biggr\rbrace\nonumber\\
&=|S(0)|^{2}\int_{-\infty}^{\infty}dt\,e^{i\omega_{0}t}\langle\sigma_{-}(t)\sigma_{+}\rangle
\label{eqn:FS}
\end{align}
where $\bar{\rho}=\lim_{T\rightarrow\infty}\rho_{T}$. Taking the complex conjugate of the expression $F(\omega_{0})$ to obtain
\begin{equation}
F^{*}(\omega_{0})=|S(0)|^{2}\int_{-\infty}^{\infty}dx\,e^{-i\omega_{0}x}\langle(\sigma_{-}(x)\sigma_{+})^{\dagger}\rangle,
\end{equation}
where
\begin{equation}
(\sigma_{-}(x)\sigma_{+})^{\dagger}=(\sigma_{-}\sigma_{+}(x)),
\end{equation}
therefore,
\begin{equation}
F^{*}(\omega_{0})=|S(0)|^{2}\int_{-\infty}^{\infty}dx\,e^{-i\omega_{0}x}\langle\sigma_{-}\sigma_{+}(x)\rangle.
\end{equation}
Performing a change of variables $x=-t$ and for stationary conditions, we have
\begin{equation}
F^{*}(\omega_{0})=|S(0)|^{2}\int_{-\infty}^{\infty}dt\,e^{i\omega_{0}t}\langle\sigma_{-}(t)\sigma_{+}\rangle.
\label{eqn:cc for FS}
\end{equation}
By comparing the equations (\ref{eqn:cc for FS}) and (\ref{eqn:FS}) one can see that $F^{*}(\omega_{0})=F(\omega_{0})$, which means that $F(\omega_{0})$ is a real function. The Fourier transform appearing in Eq.\ (\ref{eqn:FS}) can then be written as a Laplace transform 
%$F(\omega_{0})=2\mathrm{Re}[F(s=-i\omega_{0})]$
\begin{align}
F(\omega_{0})&=|S^{*}(0)|^2 \,2\mathrm{Re}\,\int_{0}^{\infty}dt\,e^{i\omega_{0}t}\mathrm{Tr}\lbrace\sigma_{-}(t)\bar{\rho}\sigma_{+}\rbrace\nonumber\\
&=|S^{*}(0)|^2 \,2\mathrm{Re}\,\int_{0}^{\infty}dt\,e^{i\omega_{0}t}\mathrm{Tr}\lbrace\sigma_{-}\Omega(t)\rbrace,
\label{eqn:fluorescence spectra}
\end{align}
where we used the cyclicity of trace in the last step, a factor of two comes from the change in the interval of integration and the operator $\Omega(t)$ is given by the expression\footnote{Since $\sigma_{+}$ and $\sigma_{-}$ are operators in the system Hilbert space and $[H_{D},H]=0$, the evolution of $\Omega(t)$ is determined by the Hamiltonian of the emitting system and radiation field, $H$, in the absence of the detector.}
\begin{equation}
\Omega(t)=e^{-iHt}\bar{\rho}\sigma_{+}e^{iHt}. 
\label{eqn:psedo density matrix}
\end{equation}
The above operator contains all the information needed for the fluorescence spectrum but it is defined in the Hilbert space of the entire world, i.e.\ the system and the reservoir.  In a sense it contains too much information so we would like to find a new operator defined only in the system Hilbert space. For this purpose, we introduce a projection method which will be discussed in the next chapter.
\section{Nakajima-Zwanzig generalized master equation} \label{sec:five}
In order to compute the spectrum given by equation (\ref{eqn:fluorescence spectra}), we write the equation for the dynamics of the operator $\Omega(t)$ and find its Laplace transform. 
Equation (\ref{eqn:fluorescence spectra}) is analogous to the expression for the single time expectation value,
\begin{equation}
\langle\sigma_{-}(t)\rangle=\mathrm{Tr}\lbrace\sigma_{-}\rho(t)\rbrace,
\label{eqn:single time expectation value}
\end{equation}
with $\rho(t)$ replaced by $\Omega(t)$  \citep{swan81}. Thus, we first write an equation of motion for the dynamics of the reduced density matrix, then the equation for the dynamics of $\Omega(t)$ can be derived in the same way.
The radiation field and system are decoupled for times $t<t_{0}$ and are prepared independently in the states described by density matrices $\rho_{R}(0)$ and $\rho_{S}(0)$, respectively. At time $t=t_{0}$, when the radiation field and system are brought into contact, the state of the entire system is described by the full density matrix $\rho(t_{0})$:
\begin{equation}
\rho(t_{0})=\rho_{S}(t_{0})\otimes\rho_{R}(t_{0}).
\label{eqn:factorised initial density matrix}
\end{equation} 
To evaluate the dynamics of the reduced
density operator, we introduce a projection superoperator $P$,
defined by its action on an arbitrary operator $\mathcal{O}$: $P\mathcal{O}
=\rho_{R}(t_{0})Tr_{R} \mathcal{O}$. $P$ is chosen to preserve all system expectation
values: 
\begin{align}
\langle \mathcal{O}_{S}\rangle(t)&=\mathrm{Tr}\lbrace\mathcal{O}_{S}\rho(t)\rbrace\nonumber\\
&=\mathrm{Tr}\lbrace \mathcal{O}_{S}P\rho(t)\rbrace\\
&=\mathrm{Tr}_{S}\mathrm{Tr}_{R}\lbrace\mathcal{O}_{S}\otimes\mathbb{1}_{R}\rbrace\rho_{R}(t_{0})\mathrm{Tr}_{R}\rho(t)\nonumber\\
&=\mathrm{Tr}_{S}\mathcal{O}_{S}\rho_{S}(t)\mathrm{Tr}_{R}\mathbb{1}_{R}\otimes\rho_{R}(t_{0})\nonumber\\
&=\mathrm{Tr}_{S}\lbrace\mathcal{O}_{S}\rho_{S}(t)\rbrace
\end{align}
and
satisfies $P^2=P$. For factorized initial conditions [Eq. (\ref{eqn:factorised initial density matrix})],
$P\rho(t_{0})=\rho(t_{0})$, which is a sufficient condition to rewrite the
von-Neumann equation 
\begin{equation}
\dot{\rho}(t)=-i[H,\rho(t)]=-iL\rho(t)
\label{eqn:von-Neumann equation for density matrix}
\end{equation}  
in the form of the
exact Nakajima-Zwanzig generalized master equation
(GME) \citep{fick}, where $L$ is the full Liouvillian, defined as $L_{\alpha}\mathcal{O}=[H_{\alpha},\mathcal{O}]$ and $\alpha=S,R,0,SR$.
Multiplying equation (\ref{eqn:von-Neumann equation for density matrix}) by $P$ on both sides, we get
\begin{equation}
P\dot{\rho}(t)=-iPL\rho(t).
\end{equation}
Introducing the complement of $P$: $Q=\mathbb{1}-P$ and using $P+Q=\mathbb{1}$ we obtain
\begin{equation}
P\dot{\rho}(t)=-iPLP\rho(t)-iPLQ\rho(t).
\label{eqn:P part of density matrix}
\end{equation}
To write the above equation in terms of $P\rho(t)$ alone, we multiply equation (\ref{eqn:von-Neumann equation for density matrix}) by $Q$, to get
\begin{equation}
Q\dot{\rho}(t)=-iQLP\rho(t)-iQLQ\rho(t).
\end{equation}
We solve the above equation for $Q\rho(t)$ using the separation of variables method
\begin{equation}
Q\dot{\rho}(t)+iQLQ\rho(t)=-iQLP\rho(t).
\end{equation}
Multiplying the above equation by $e^{iQLt}$ on both sides gives
\begin{align}
e^{iQLt}Q\dot{\rho}(t)+ie^{iQLt}QLQ\rho(t)&=-ie^{iQLt}QLP\rho(t)\nonumber\\
\frac{d}{dt}e^{iQLt}Q\rho(t)&=-ie^{iQLt}QLP\rho(t).
\end{align}
After integrating the above equation and assuming $Q\rho(t_{0})=0$, i.e.,\ assuming that the so-called final part of $\rho(t_{0})$ is zero, we obtain
\begin{equation}
Q\rho(t)=-i\int_{t_{0}}^{t}dt'e^{-iQL(t-t')}QLP\rho(t').
\label{eqn:Q part of density matrix}
\end{equation}
Substituting equation (\ref{eqn:Q part of density matrix}) into equation (\ref{eqn:P part of density matrix}), we obtain the standard form of the Nakajima-Zwanzig generalized master equation \citep{fick}
\begin{equation}
P\dot{\rho}(t)=-iPLP\rho(t)-i\int_{t_{0}}^{t}dt'\ \Sigma(t-t')P\rho(t'),
\label{eqn:GME for density matrix}
\end{equation}
\begin{equation}
\Sigma(t)=-iPLQ\ e^{-iQLt}QLP,
\label{eqn:self-energy superoperator}
\end{equation}
where $\Sigma(t)$ is the self-energy superoperator. We can derive an equation of motion for $\Omega(t)$ analogous to the equation for $\rho(t)$ [Eq. (\ref{eqn:GME for density matrix})].  However, an additional term appears because $Q\Omega(0)=Q\bar{\rho}\sigma_+ \neq 0$, see Eq.\ (\ref{eqn:Q part of density matrix}).  The resulting GME for $P \Omega (t)$ is then
\begin{equation}
P\dot{\Omega}(t)=-iPLP\Omega(t)-i\int_{0}^{t}dt'\ \Sigma(t-t')\Omega(t')-iPLQe^{-iQLt}Q\Omega(0),
\label{eqn:GME for pseudo density matrix}
\end{equation}
where $\Sigma(t)$ is defined in Eq.\ (\ref{eqn:self-energy superoperator}) and the last term in the above equation contains $Q\Omega(0)$, i.e.\ the final part of $\Omega(0)$ is non-zero. This last term accounts for conditions that accumulate between the system and radiation modes in the time interval $t\in[ t_{0},0]$ for $t_{0}<0$. The long-time average value, is defined as
\begin{equation}
\bar{\rho}=\lim_{T\rightarrow \infty}\frac{1}{T}\int_{0}^{T}dt\rho(t)=\lim_{s\rightarrow 0}s\rho(s).
\label{eqn:stationary limit density matrix}
\end{equation} 
Here, the Laplace transform is defined as $F(s)=\int_{t_{0}}^{\infty}dt\, e^{-st}f(t)$. Inserting this definition into Eq. (\ref{eqn:Q part of density matrix}), we find (assuming that $\frac{1}{0^{+}+iQL}$ exists),
\begin{equation}
Q\bar{\rho}=-i\frac{1}{0^{+}+iQL}QLP\bar{\rho}.
\end{equation}
Here, $0^+$ is a positive infinitesimal. We want to write $Q\Omega(0)=Q\bar{\rho}\sigma_{+}$ in terms of $P\bar{\rho}\sigma_{+}$. To this end, substituting for $Q\bar{\rho}$ in equation (\ref{eqn:GME for pseudo density matrix}), we have
\begin{equation}
P\dot{\Omega}(t)=-iPLP\Omega(t)-i\int_{0}^{t}dt'\ \Sigma(t-t')\Omega(t')-PLQe^{-iQLt}\left(\frac{1}{0^{+}+iQL}\right)QLP\bar{\rho}\sigma_{+}.
\label{eqn:GME for pseudo density matrix with irrelevant part in time domain}
\end{equation}
The above equation is identical to equation (\ref{eqn:GME for density matrix}) except for the last part, the  so-called final part, which is expressed as
\begin{equation}
\Phi(t)=-PLQe^{-iQLt}\left(\frac{1}{0^{+}+iQL}\right)QLP\bar{\rho}\sigma_{+}.
\label{eqn:irrelevant part in time domain}
\end{equation}
The expression for the fluorescence spectrum given in equation (\ref{eqn:fluorescence spectra}) is written in terms of the trace of the system operator $\sigma_{-}$ as $\mathrm{Tr}\lbrace\sigma_{-}\Omega(t)\rbrace$.

Using the properties of the projection superoperator, we have
\begin{equation}
\mathrm{Tr}\lbrace\sigma_{-}\Omega(t)\rbrace=\mathrm{Tr}\lbrace \sigma_{-}\mathrm{P}\Omega(t)\rbrace
=\mathrm{Tr}_{S}\lbrace\sigma_{-}\Omega_{S}(t)\rbrace,
\end{equation}
where in the last step we have defined 
\begin{equation}
\Omega_{S}(t)=\mathrm{Tr}_{R}\lbrace\Omega(t)\rbrace.
\label{eqn:reduced operator in time}
\end{equation}
Therefore, the expression for the fluorescence spectrum given in Eq.\ (\ref{eqn:fluorescence spectra}) takes the form
\begin{equation}
F(\omega_{0})=2|S^{*}(0)|^2\mathrm{Re}\int_{0}^{\infty}dt\,e^{i\omega_{0} t}\mathrm{Tr}_{S}\lbrace\sigma_{-}\Omega_{S}(t)\rbrace.
\end{equation}
For completeness, we substitute the expression for the Laplace transform $P\Omega(s)$ in the equation for the fluorescence spectrum to obtain an expression for the lineshape function as
\begin{equation}
F(\omega_{0})=2|S^{*}(0)|^{2}\mathrm{Re}\biggl[\mathrm{Tr}\biggl\lbrace\sigma_{-}\frac{1}{s+iPLP+i\Sigma(s)}\left(\mathbb{1}+\Phi(s)\right)P\bar{\rho}\sigma_{+}\biggr\rbrace\biggr]_{s=-i\omega_{0}}
\label{eqn:GME for pseudo density matrix in laplace domain}
\end{equation}
where the Laplace transforms of the self-energy and the \emph{final part} are given, respectively, below  
\begin{equation}
\Sigma(s=-i\omega_{0})=-iPLQ\frac{1}{-i\omega_{0}+iQL}QLP
\label{eqn:self-energy in laplace domain}
\end{equation}
\begin{equation}
\Phi(s=-i\omega_{0})=-PLQ\left(\frac{1}{-i\omega_{0}+iQL}\right)\ \left(\frac{1}{0^{+}+iQL}\right)QLP,
\label{eqn:irrelevant part in laplace domain}
\end{equation}
\begin{equation}
\bar{\rho}=\lim_{s\rightarrow 0}s\frac{1}{s+iPLP+i\Sigma(s)}P\rho(t_{0}).
\label{eqn:stationary limit density matrix in laplace domain}
\end{equation}
We have derived a formula which is valid for studying the fluorescence spectrum of a general system undergoing non-Markovian dynamics.
\section{Results and discussion} \label{sec:six}
In this section, we analyze and discuss the results obtained in the previous sections. The self-energy and final terms given by Eqs.\ (\ref{eqn:self-energy in laplace domain}) and (\ref{eqn:irrelevant part in laplace domain}), respectively can be expanded to the all powers of perturbation/interaction (e.g. nuclear spins, phonons etc) present in the problem. Similarly, the stationary density matrix in Eq.\ (\ref{eqn:stationary limit density matrix in laplace domain}) can be found after expanding the self-energy superoperator in the powers of perturbation Liouvillian. After solving for a specific interaction, the Eq.\ (\ref{eqn:GME for pseudo density matrix in laplace domain}) will give rise to a final expression for the lineshape function of a multi-level system within non-Markovian interaction \citep{abhi2018}. 
\section{Conclusion} \label{sec:seven}
In the present paper, we have given a general analytical formula for the dynamics (fluorescence lineshape) of a multi-level system interacting with its environment via a non-Markovian interaction. We have also shown that quantum regression theorem can not be used to describe the dynamics of two-time correlation because of a non-zero final term. The self-energy and final term superoperators can be expanded in all powers of perturbation without applying the Born-approximation in terms of coupling to the environment. We have tried to keep the formulae as general as possible and not made any assumptions about the system or the environment, so that this theory can be applied to any quantum-optical system (not limited to a two-level system) or the environment (nuclear spins or phonons). Furthermore, for a Markovian type interaction above formulae can be reduced and used to study the systems with Markovian interactions and vanishing final term.
\section*{Acknowledgments}
AK acknowledge financial support from the Icelandic Research Fund RANNIS and CIFAR, Canada. AK thanks  Sigurdur I. Erlingsson and Bill Coish for the useful discussions and feedback.


\begin{thebibliography}{22}
\providecommand{\natexlab}[1]{#1}
\providecommand{\url}[1]{\texttt{#1}}
\expandafter\ifx\csname urlstyle\endcsname\relax
  \providecommand{\doi}[1]{doi: #1}\else
  \providecommand{\doi}{doi: \begingroup \urlstyle{rm}\Url}\fi

\bibitem[Peng et~al.(2019)Peng, Yang, Wu, and Li]{peng2019}
Ze-an Peng, Guo-qing Yang, Qing-lin Wu, and Gao-xiang Li.
\newblock Filtered strong quantum correlation of resonance fluorescence from a
  two-atom radiating system with interatomic coherence.
\newblock \emph{Phys. Rev. A}, 99:\penalty0 033819, Mar 2019.

\bibitem[Konthasinghe et~al.(2019)Konthasinghe, Chakraborty, Mathur, Qiu,
  Mukherjee, Fuchs, and Vamivakas]{Konthasinghe2019}
Kumarasiri Konthasinghe, Chitraleema Chakraborty, Nikhil Mathur, Liangyu Qiu,
  Arunabh Mukherjee, Gregory~D. Fuchs, and A.~Nick Vamivakas.
\newblock Rabi oscillations and resonance fluorescence from a single hexagonal
  boron nitride quantum emitter.
\newblock \emph{Optica}, 6\penalty0 (5):\penalty0 542--548, May 2019.

\bibitem[Schöll et~al.(2019)Schöll, Hanschke, Schweickert, Zeuner, Reindl,
  Covre~da Silva, Lettner, Trotta, Finley, Müller, Rastelli, Zwiller, and
  Jöns]{scholl2019}
Eva Schöll, Lukas Hanschke, Lucas Schweickert, Katharina~D. Zeuner, Marcus
  Reindl, Saimon~Filipe Covre~da Silva, Thomas Lettner, Rinaldo Trotta,
  Jonathan~J. Finley, Kai Müller, Armando Rastelli, Val Zwiller, and Klaus~D.
  Jöns.
\newblock Resonance fluorescence of gaas quantum dots with near-unity photon
  indistinguishability.
\newblock \emph{Nano Letters}, 19\penalty0 (4):\penalty0 2404--2410, 2019.

\bibitem[Yang and An(2016)]{yang2016}
Chun-Jie Yang and Jun-Hong An.
\newblock Resonance fluorescence beyond the dipole approximation of a quantum
  dot in a plasmonic nanostructure.
\newblock \emph{Phys. Rev. A}, 93:\penalty0 053803, May 2016.

\bibitem[Ulrich et~al.(2011)Ulrich, Ates, Reitzenstein, L\"offler, Forchel, and
  Michler]{ulrich11prl}
S.~M. Ulrich, S.~Ates, S.~Reitzenstein, A.~L\"offler, A.~Forchel, and
  P.~Michler.
\newblock Dephasing of triplet-sideband optical emission of a resonantly driven
  $\mathrm{InAs}/\mathrm{GaAs}$ quantum dot inside a microcavity.
\newblock \emph{Phys. Rev. Lett.}, 106:\penalty0 247402, Jun 2011.

\bibitem[Mollow(1969)]{mollow69}
B.~R. Mollow.
\newblock Power spectrum of light scattered by two-level systems.
\newblock \emph{Phys. Rev.}, 188:\penalty0 1969--1975, Dec 1969.
\newblock \doi{10.1103/PhysRev.188.1969}.

\bibitem[Swain(1981)]{swan81}
S~Swain.
\newblock Master equation derivation of quantum regression theorem.
\newblock \emph{Journal of Physics A: Mathematical and General}, 14\penalty0
  (10):\penalty0 2577, 1981.

\bibitem[Lax(2000)]{lax00}
Melvin Lax.
\newblock The lax–onsager regression `theorem' revisited.
\newblock \emph{Optics Communications}, 179\penalty0 (1–6):\penalty0 463 --
  476, 2000.
\newblock ISSN 0030-4018.

\bibitem[Lax(1968)]{lax68}
Melvin Lax.
\newblock Quantum noise. xi. multitime correspondence between quantum and
  classical stochastic processes.
\newblock \emph{Phys. Rev.}, 172:\penalty0 350--361, Aug 1968.

\bibitem[Coish and Loss(2004)]{bill04}
W.~A. Coish and Daniel Loss.
\newblock Hyperfine interaction in a quantum dot: Non-markovian electron spin
  dynamics.
\newblock \emph{Phys. Rev. B}, 70:\penalty0 195340, Nov 2004.

\bibitem[Kumar(2018)]{abhi2018}
Abhishek Kumar.
\newblock Theory of non-markovian dynamics in resonance fluorescence spectrum.
\newblock \emph{Optical and Quantum Electronics}, 50\penalty0 (8):\penalty0
  317, 2018.

\bibitem[Krummheuer et~al.(2002)Krummheuer, Axt, and Kuhn]{krummheuer02}
B.~Krummheuer, V.~M. Axt, and T.~Kuhn.
\newblock Theory of pure dephasing and the resulting absorption line shape in
  semiconductor quantum dots.
\newblock \emph{Phys. Rev. B}, 65:\penalty0 195313, May 2002.

\bibitem[Vagov et~al.(2014)Vagov, Gl\"assl, Croitoru, Axt, and Kuhn]{vagov14}
A.~Vagov, M.~Gl\"assl, M.~D. Croitoru, V.~M. Axt, and T.~Kuhn.
\newblock Competition between pure dephasing and photon losses in the dynamics
  of a dot-cavity system.
\newblock \emph{Phys. Rev. B}, 90:\penalty0 075309, Aug 2014.

\bibitem[Weiler et~al.(2012)Weiler, Ulhaq, Ulrich, Richter, Jetter, Michler,
  Roy, and Hughes]{weiler12}
S.~Weiler, A.~Ulhaq, S.~M. Ulrich, D.~Richter, M.~Jetter, P.~Michler, C.~Roy,
  and S.~Hughes.
\newblock Phonon-assisted incoherent excitation of a quantum dot and its
  emission properties.
\newblock \emph{Phys. Rev. B}, 86:\penalty0 241304, Dec 2012.

\bibitem[Roy and Hughes(2012)]{hughes12prb}
C.~Roy and S.~Hughes.
\newblock Polaron master equation theory of the quantum-dot mollow triplet in a
  semiconductor cavity-qed system.
\newblock \emph{Phys. Rev. B}, 85:\penalty0 115309, Mar 2012.

\bibitem[Fick and Sauermann(1990)]{fick}
E.~Fick and G.~Sauermann.
\newblock \emph{The Quantum Statistics of Dynamic Processes}.
\newblock Springer-Verlag, Berlin, 1990.

\bibitem[de~Vega and Alonso(2008)]{vega08}
In\'es de~Vega and Daniel Alonso.
\newblock Emission spectra of atoms with non-markovian interaction:
  Fluorescence in a photonic crystal.
\newblock \emph{Phys. Rev. A}, 77:\penalty0 043836, Apr 2008.

\bibitem[Guarnieri et~al.(2014)Guarnieri, Smirne, and Vacchini]{guarnieri2014}
Giacomo Guarnieri, Andrea Smirne, and Bassano Vacchini.
\newblock Quantum regression theorem and non-markovianity of quantum dynamics.
\newblock \emph{Phys. Rev. A}, 90:\penalty0 022110, Aug 2014.

\bibitem[Ford and O'Connell(1996)]{ford96}
G.~W. Ford and R.~F. O'Connell.
\newblock There is no quantum regression theorem.
\newblock \emph{Phys. Rev. Lett.}, 77:\penalty0 798--801, Jul 1996.

\bibitem[Scully and Zubairy(1997)]{scully}
Marlan~O. Scully and M.~Suhail Zubairy.
\newblock \emph{Quantum Optics}.
\newblock Cambridge University Press, Cambridge, 1997.

\bibitem[Cohen-Tannoudji and Grynberg(2004)]{cohen}
Jaques Dupont-Roc Cohen-Tannoudji and Gilbert Grynberg.
\newblock \emph{Atom-Photon Interactions}.
\newblock Wiley-VCH, Berlin, 2004.

\bibitem[Bruus and Flensberg(2004)]{bruus}
Henrik Bruus and Karsten Flensberg.
\newblock \emph{Many-body Quantum Theory in Condensed Matter Physics}.
\newblock Oxford University Press, Cambridge, 2004.

\end{thebibliography}
\end{document}